\documentclass[12pt]{article}

\usepackage{hyperref}
\usepackage{graphicx}
\usepackage{amssymb}
\usepackage{amsmath}
\usepackage{mathrsfs}

\def\bea{\begin{eqnarray}}
\def\eea{\end{eqnarray}}
\def\nn{\nonumber}

\begin{document}
\title{Luminosity and Crab Waist Collision Studies}
\author{Wanwei Wu\footnote{Email: wwu1@go.olemiss.edu}  \, and Don Summers\footnote{Email: summers@phy.olemiss.edu}\\[1ex] \small{Department of Physics and Astronomy, University of Mississippi, University, MS 38677}}
\date{}
\maketitle


\section*{\centering Abstract} 

\hspace*{\parindent}In high energy physics, the luminosity is one useful value to characterize the performance of a particle collider. To gain more available data, we need to maximize the luminosity in most collider experiments. With the discussions of tune shift involved the beam dynamics and a recently proposed ``crabbed waist" scheme of beam-beam 
collisions, we present some qualitative analysis to increase the luminosity. In addition, beam-beam tune shifts and luminosities of  $e^{+}e^{-}$, $pp$/$p\bar{p}$, and $\mu^{+}\mu^{-}$ colliders are discussed.

\section{Introduction}
\subsection{Luminosity}
\hspace*{\parindent}In high energy physics, the beam energy and the luminosity are usually the two most important parameters that quantify the performance of particle colliders. The former one is necessary to provide available energy for production of new effects, i.e., a new massive particle; while the latter one measures the ability of a particle accelerator to produce the required number of interactions, which is especially important when there are only rare events with a small production cross section $\sigma$ (the total area of overlap of two colliding particles). 

\begin{figure}
\begin{center}
\includegraphics[width=10cm, height=4.5cm]{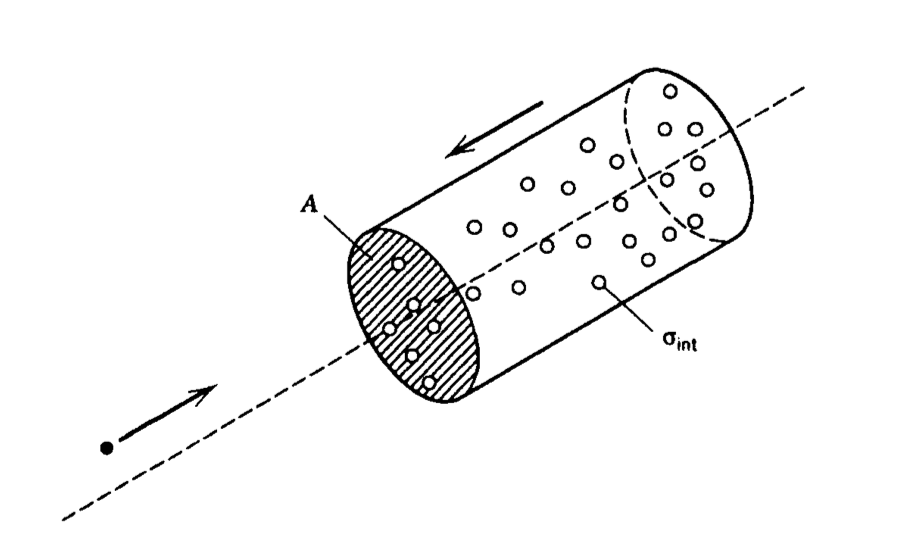}
\end{center}
\caption{Collision of a single test particle from one beam with a particle bunch of the other beam \cite{Fluminosity}}
\label{Fcollision}
\end{figure}

The particle accelerators cannot produce continuous particle beams but beams consisting of a sequence of ``bunches" of particles because of the time varying fields used to produce acceleration. Therefore, it is two such beam bunches that are brought into collision in fact. Suppose one bunch of $N_{1}$ particles moving in one direction collides head-on with a bunch of $N_{2}$ particles in the opposite direction. Both of them have cross-sectional area $A$. As shown in Fig.~\ref{Fcollision}, any single particle in one bunch ($N_{1}$) has a chance of a fraction of the area $N_{2}\sigma_{int}/A$ to ``meet" a particle in the other bunch ($N_{2}$). The number of interactions per passage of two such bunches is then $N_{1}N_{2}\sigma_{int}/A$. Hence, the interaction rate is given by
\bea
R&=&f\frac{N_{1}N_{2}}{A}\sigma_{int},
\eea
where f is the frequency of bunch collisions.

The luminosity, $\mathscr{L}$, is defined as the interaction rate per unite cross section:
\bea
\mathscr{L}&=&\frac{R}{\sigma_{int}}=f\frac{N_{1}N_{2}}{A}.
\eea
For a ``Round Gaussian" particle distribution, the luminosity can be written as 
\bea
\mathscr{L}&=&f\frac{N_{1}N_{2}}{4\pi\sigma_{x}\sigma_{y}},
\label{Elhead}
\eea 
where $\sigma_{x}$ and $\sigma_{y}$ characterize the Gaussian transverse beam profiles in the horizontal (bend) and vertical directions and present the sizes of the beam. Luminosities are often expressed in \textit{cgs} units: $cm^{-2}s^{-1}$. A glance at the luminosity formula reveals that to raise luminosity one must increase the collision frequency, bunch intensity and lower the beam cross sectional area.

\subsection{The Beam-Beam Tune Shift}

\hspace*{\parindent}The number of betatron oscillations (bounded oscillatory motion about the design trajectory corresponding to the transverse stability) per turn in a synchrotron is called \textit{tune} and given by \cite{Fluminosity01}.
\bea
\nu &=& \frac{1}{2\pi}\oint\frac{ds}{\beta(s)},
\eea
where the independent variable $s$ is path length along the design trajectory and the quantity $\beta(s)$ is usually referred to as the amplitude function. In a colliding beam accelerator, each time the beams cross each other, the particles in one beam feel the electric and magnetic forces due to the particles in the other beam. Since the particles in these two beams have opposite velocity directions, the electric and magnetic forces do not cancel but rather add, creating a net defocussing force. In other words, the particles of one bunch see the other bunch as a nonlinear lens in a bunch-bunch collision. For particles undergoing infinitesimal betatron oscillations in a highly relativistic Gaussian beam, the net force would be 
\bea
F&=&\frac{e^{2}N}{2\pi\epsilon_{0}\sigma^{2}}r,
\eea
where $N$ is the number of particles of each bunch ($N_{1}=N_{2}=N$), $r$ is the radius of the accelerating orbit, and $\sigma^{2}=\sigma_{x}\sigma_{y}$ ($\sigma_{x}$ and $\sigma_{y}$ are the horizontal and vertical beam sizes). The tune shift experienced by the particle would be
\bea
\Delta\nu&=&\frac{1}{4\pi}\frac{1}{pc}\frac{e^2}{2\pi\epsilon_{0}}\oint\frac{N\beta(s)}{\sigma^{2}(s)}ds \nn \\
&=&\frac{r_{0}}{2\epsilon_{N}}\times \frac{1}{2}\int N ds.
\eea
Here, $r_{0}=e^{2}/(4\pi\epsilon_{0}mc^{2})$ is the classical radius of the particle. For the proton, $r_{p}=r_{0}=1.53\times 10^{-18}$ meters; 
for the muon
$r_{\mu}=r_{0}=1.36\times 10^{-17}$ meters; and
for the electron, $r_{e}=r_{0}=2.82 \times 10^{-15}$ meters. $\epsilon_{N}=\pi\sigma^{2}(\gamma \beta)/\beta(s) =\pi (\gamma\beta) \sqrt{\epsilon_{x}\epsilon_{y}}$ 
is the normalized emittance ($\epsilon_{x}$ and $\epsilon_{y}$ are the horizontal and vertical emittance). The definition of $\epsilon_{N}$ may be different by a factor of ``$\pi$" \cite{mmrad}. In terms of the total number of particles in a bunch, $N$, the beam-beam tune shift per collision becomes \cite{Fluminosity01}
\bea
\Delta\nu =   \xi = \frac{Nr_{0}}{4\epsilon_{N}}.
\label{Ets}
\eea

\subsection{The ``Crab Waist" Scheme}

\hspace*{\parindent}A recently proposed ``crabbed waist" scheme of beam-beam collisions \cite{LPCW} can substantially increase the luminosity of a collider since it combines several potentially advantageous ideas. To discuss such potential advantages, let's try to describe the crabbed waist concept first.

\begin{figure}
\begin{center}
\includegraphics[width=7cm, height=3cm]{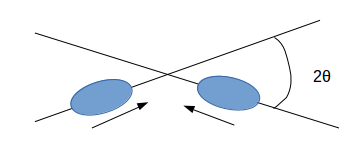}
\includegraphics[width=7cm, height=3cm]{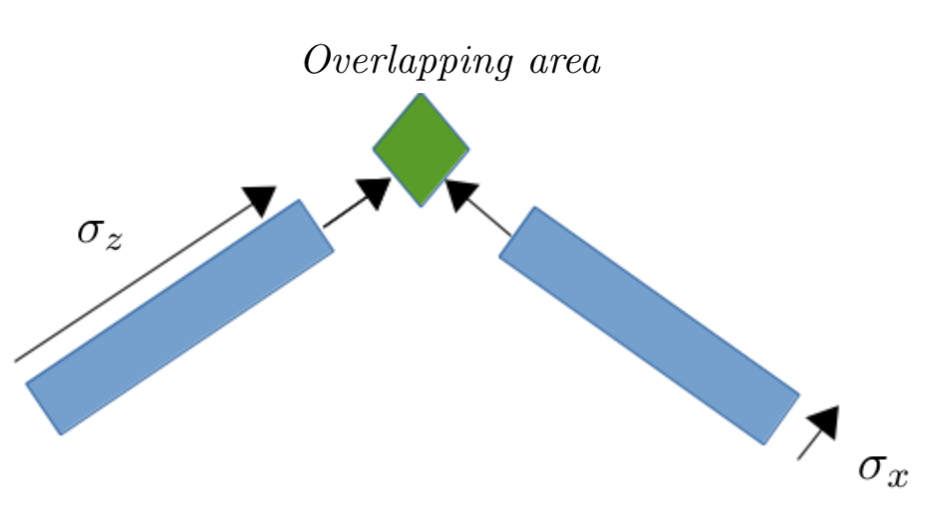}
\end{center}
\caption{Scheme of Collision with a Crossing Angle}
\label{Fcrossangle}
\end{figure}

\begin{figure}
\begin{center}
\includegraphics[width=13cm, height=3cm]{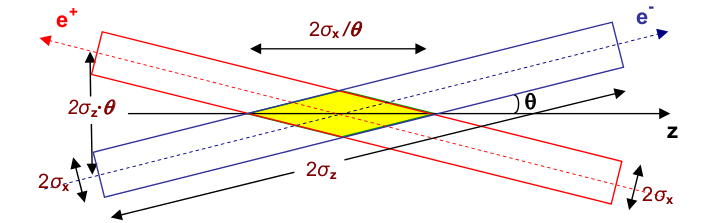}
\end{center}
\caption{The ``Crabbed Waist" Collision Scheme \cite{CWscheme}}
\label{FCW}
\end{figure}

In Fig.~\ref{Fcrossangle}, the crossing angle $2\theta$ (the left graph) and the overlapping area (the right graph) are shown. For collisions under a crossing angle $2\theta$, the luminosity $\mathscr{L}$, the horizontal $\xi_{x}$ and vertical $\xi_{y}$ tune shift scale as \cite{CWscheme, CWmzpr}
\begin{equation}
\mathscr{L}\propto\frac{N\xi_{y}}{\beta_{y}^{*}}\propto \frac{1}{\sqrt{\beta_{y}^{*}}};\  \xi_{y}\propto \frac{N\beta_{y}^{*}}{\sigma_x \sigma_{y}\cdot \sqrt{1+\phi^2}};\  \xi_{x}\propto \frac{N}{\epsilon_{x}\cdot \sqrt{1+\phi^2}}.
\end{equation}
Here, $\phi$ is the Piwinski angle, defined as
\bea
\phi&=&\frac{\sigma_{z}}{\sigma_{x}}\tan\theta \approx\frac{\sigma_{z}}{\sigma_{x}}\theta,
\eea
where $\sigma_{x,y,z}$ are the bunch beam sizes, $\beta_{y}^{*}$ is the vertical beta-function at the interaction point (IP), $\xi_{y}$ is the vertical beam-beam tune shift, and $\epsilon_{x}$ is the horizontal betatron emittance, as shown in Fig.~\ref{FCW}. The ``crabbed waist" collision means that the crossing angle $2\theta\gg \sigma_{x}/\sigma_{z}$, in contrast, the ``head-on" collision means that $2\theta\ll \sigma_{x}/\sigma_{z}$.

High luminosity requires short bunches with a very small vertical beta-function $\beta_{y}^{*}$ at the IP and a high beam intensity $I$ with small vertical emittance $\epsilon_{y}$.
Large horizontal beam size $\sigma_{x}$ (see Fig.~\ref{Fcrossangle}) and large horizontal emittance $\epsilon_{x}$ can be tolerated.
Though short bunches are usually hard to make,  with large Piwinski angle $\phi$ the overlapping area becomes much smaller than $\sigma_z$, allowing significant $\beta_y$ decrease and a very significant potential gain in the luminosity \cite{CWscheme}. 
The large Piwinski angle increases luminosity by allowing a small $\beta^*_y$, but  decreases the luminosity by effectively decreasing $N.$
However, the beam-beam tune shift is now small and a large gain in luminosity may be possible by raising the beam-beam tune shift to its limit.

\section{Electron-Positron Collider}

\hspace*{\parindent}Due to the synchrotron radiation, the energy loss by one electron in a circular orbit is
\bea
\delta E&=&\frac{8.85\times 10^{-5}E^{4}(GeV)}{R(meter)}.
\eea
Because an electron has a small mass, the transverse beam size is  naturally damped by the synchrotron radiation.
Electron-positron colliders are typically synchrotron wall power limited, so more particles cannot be added.
One can take advantage of the crab waist crossing by lowering emittance.

\subsection{``Head-on" Collision}

\hspace*{\parindent}In head-on collisions, the vertical beam-beam tune shift \cite{Htuneshift, BookPAP} is 
\bea
\xi_{y}&=&\frac{Nr_{e}\beta_{y}^{*}}{2\pi\gamma\sigma_{y}(\sigma_{x}+\sigma_{y})}.
\label{Exiy}
\eea
Also, the horizontal beam-beam tune shift $\xi_{x}$ is given by
\bea
\xi_{x}&=&\frac{Nr_{e}\beta_{x}^{*}}{2\pi\gamma\sigma_{x}(\sigma_{x}+\sigma_{y})}.
\eea
For ``round beam cross sections", i.e., $\sigma_{x}\approx\sigma_{y}$, using the definition of normalized emittance $\epsilon_{N}=\pi\sigma^2(\gamma v/c)/\beta^{*}$, at a relativistic high energy ($v\to c$) we will have 
\bea
\xi_{y}&=&\frac{Nr_{e}}{4\epsilon_{N}}.
\eea

For ``elliptical beam cross sections", i.e., $\sigma_{x}\neq\sigma_{y}$, with the assumption that the elliptical beams have the same cross sectional area and charge density as the round beams \cite{Brown1987}, we have $\sigma^{2}=\sigma_{x}\sigma_{y}$ and the normalized emittance $\epsilon_{N}=\pi\gamma\sqrt{\epsilon_{x}\epsilon_{y}}$, where $\epsilon_{x}=\frac{\sigma_{x}^{2}}{\beta_{x}^{*}}$ and $\epsilon_{y}=\frac{\sigma_{y}^{2}}{\beta_{y}^{*}}$.
The Eq. (\ref{Exiy}) of vertical beam-beam tune shift then becomes
\bea
\xi_{y}&=&\frac{Nr_{e}\beta_{y}^{*}}{2\pi\gamma\sqrt{\beta_{y}^{*}\epsilon_{y}}(\sqrt{\beta_{x}^{*}\epsilon_{x}}+\sqrt{\beta_{y}^{*}\epsilon_{y}})}\nn\\
&=&\frac{Nr_e}{2\epsilon_{N}}\cdot\frac{1}{\sqrt{\frac{\beta_{x}^{*}}{\beta_{y}^{*}}}+\sqrt{\frac{\epsilon_{y}}{\epsilon_{x}}}}
\eea
As the same, the horizontal beam-beam tune shift becomes
\bea
\xi_{x}&=&\frac{Nr_{e}\beta_{x}^{*}}{2\pi\gamma\sqrt{\beta_{x}^{*}\epsilon_{x}}(\sqrt{\beta_{x}^{*}\epsilon_{x}}+\sqrt{\beta_{y}^{*}\epsilon_{y}})}\nn\\
&=&\frac{Nr_e}{2\epsilon_{N}}\cdot\frac{1}{\sqrt{\frac{\epsilon_{x}}{\epsilon_{y}}}+\sqrt{\frac{\beta_{y}^{*}}{\beta_{x}^{*}}}}
\eea

For the beam cross sections $\sigma_{y}\ll\sigma_{x}$, $\xi_{y}\approx \frac{Nr_{e}\beta_{y}^{*}}{2\pi\gamma\sigma_{x}\sigma_{y}}$, we will have 
\bea
\xi_{y}&=&\frac{Nr_{e}}{2\epsilon_{N}}.
\label{Ehts}
\eea

\begin{table}[ht] 
\caption{LEP Beam Parameters at Three Different Energies \cite{LEP}}\smallskip
\centering   
\tabcolsep=1.2mm   
\begin{tabular}{l c c c c c c c c c c} 
\hline\hline\
$E$ & $N$ & $k_{b}$ & $\mathscr{L}$ & $Q_s$ & $Q$ & $\beta^*$ & $\epsilon$ & $\sigma$ & $\xi$\\ 
$(GeV)$ & ($\times10^{11}$) & & ($cm^{-1}s^{-2}$) & & & (m) & (nm) & ($\mu m$)& &\\ [0.5ex] 
\hline    
45.6 & 1.18 & 8 & $1.51\times 10^{31}$ & 0.065 & 90.31 & 2.0 & 19.3 & 197 & 0.030\\
& & & & & 76.17 & 0.05 & 0.23 & 3.4 & 0.044\\ [0.5ex]
65 & 2.20 & 4 & $2.11\times 10^{31}$ & 0.076 & 90.26 & 2.5 & 24.3 & 247 & 0.029\\
& & & & & 76.17 & 0.05 & 0.16 & 2.8 & 0.051\\ [0.5ex]
97.8 & 4.01 & 4 & $9.73\times 10^{31}$ & 0.116 & 98.34 & 1.5 & 21.1 & 178 & 0.043\\
& & & & & 96.18 & 0.05 & 0.22 & 3.3 & 0.079\\ [0.5ex] 
\hline\hline     
\end{tabular} 
\label{LEPtable}  
\end{table} 

The luminosity then can be written as
\bea
\mathscr{L}&=&\frac{fN_{1}N_{2}}{4\pi\sigma_{x}\sigma_{y}}=\frac{fN_{1}N_{2}\gamma}{4\epsilon_{N}\beta_{y}^{*}}=\frac{fN_{1}\gamma\xi_{y}}{2r_{e}\beta_{y}^{*}}.
\eea

As an example, let's consider the LEP collider with energy $E=97.8\  GeV$ per beam. The corresponding parameters are given in Table \ref{LEPtable}. Therefore, the normalized emittance is 
\bea
\epsilon_{N}&=&\pi\times \frac{97.8\times 10^{3}}{0.511} \sqrt{21.1\times 10^{-9}\times 0.22\times 10^{-9}}=1.295 \times 10^{-3}
\eea
The beam-beam tune shift is 
\bea
\xi_{y}&=&\frac{Nr_{e}}{2\epsilon_{N}}\cdot\frac{1}{\sqrt{\frac{\beta_{x}^{*}}{\beta_{y}^{*}}}+\sqrt{\frac{\epsilon_{y}}{\epsilon_{x}}}}\nn\\
&=&\frac{4.01\times 10^{11}\times 2.82\times 10^{-15}}{2\times 1.295*10^{-3}}\times\frac{1}{\sqrt{\frac{1.5}{0.05}}+\sqrt{\frac{0.22}{21.2}}}\nn\\
&\approx&0.078 
\eea

\begin{table}[ht] 
\caption{$e^{+}e^{-}$ Collider Parameters \cite{TLyons}}\smallskip
\centering   
\tabcolsep=1.2mm   
\begin{tabular}{l c c c c} 
\hline\hline\                      
Parameter & LEP & VLLC & $Crab Waist_{200}$ & $Crab Waist_{250}$\\ [0.5ex] 
\hline                    
Circumference (m) & 26 658.9 & 233 000.0 & 233 000.0 & 233 000.0  \\   
$\beta^*_{x}$, $\beta^*_{y}$ (cm) & 150, 5 & 100, 1 & 2, .06 & 2, .06  \\ 
Luminosity $(cm^{-2}$ $sec^{-1})$ & 9.73 $\times$ $10^{31}$ & 8.8 $\times$ $10^{33}$ & 1.5$\times$ $10^{35}$ & 9.7$\times$ $10^{34}$ \\ 
Energy (GeV) & 97.8 & 200.0 & 200.0 & 250.0\\ 
$\gamma$ & 191 000 & 391 000 & 391 000 & 489 000\\
Emittances $\epsilon_x$, $\epsilon_y$ (nm) & 21.1, 0.220 & 3.09, 0.031 & .9, .0017 & .9, .00067\\
rms beam size IP $\sigma^*_x$, $\sigma^*_y$ ($\mu$m) & 178.0, 3.30 & 55.63, 0.56 & 4.25, 0.0321 & 4.25, 0.0201\\ 
Bunch intensity/I (/mA) & ${4.01 \times 10^{11}}\over 0.720$ & ${4.85 \times 10^{11}}\over 0.1$ & ${4.85 \times 10^{11}}\over 0.1$ & ${4.85 \times 10^{11}}\over 0.04$\\ 
Number of bunches per beam & 4 & 114 & 114 & 46\\ 
Total beam current (mA) & 5.76 & 22.8 & 22.8 & 9.34\\ 
Beam-beam tune shift $\xi_{x}$, $\xi_{y}$ & 0.043, 0.079 & 0.18, 0.18 & 0.034, 0.18  & 0.027, 0.23\\ 
Dipole field (T) & 0.110 & 0.0208 & 0.0208 & 0.0260\\ 
E loss / $e^{\pm}$ / turn (GeV) & 2.67 & 4.42 & 4.42 & 10.8\\ 
Bunch length (mm) & 11.0 & 6.67 &  6.67 &  6.67\\ 
Revolution frequency (kHz) & 11.245 &  1.287 & 1.287 & 1.287\\ 
Synch rad pwr (b.b.) (MW) & 14.5 & 100.7 & 100.7 & 100.7\\  [1ex]       
\hline\hline     
\end{tabular} 
\label{table}  
\end{table}

For a specific electron-positron bunch collision, the luminosity $\mathscr{L}$ can be given by
\bea
\mathscr{L}&=&\frac{1}{4er_{e}}\frac{\xi_{y}}{\beta_{y}^{*}}\gamma I,
\label{Eeeheadon}
\eea
where $r_e=e^2/(4\pi\epsilon_{0}mc^2)=2.82\times 10^{-15}\ m$ is the classical radius of the electron, $\gamma$ is the relativistic scaling factor and $I=2feN$ is the total beam current of both beams. 
Let's consider the luminosity of a Very Large Lepton Collider (VLLC) proposed in 2002. Using Eq. (\ref{Eeeheadon}) and the parameters proposed in Table \ref{table}\  \cite{TLyons, VLLC}, we can get
\bea
\mathscr{L}_{VLLC}&=&\frac{1}{4*1.602\times10^{-19}C*2.82\times10^{-15}m}\frac{0.18}{1cm}\times 391000\times 22.8mA\nn\\
&=&8.8\times10^{33}\ cm^{-2}s^{-1}
\label{Evllc}
\eea

\subsection{``Crab Waist" Collision}

\hspace*{\parindent}In the ``crabbed waist" scheme \cite{PRLMZ}, the vertical beam-beam tune shift is (a factor of $1/2$ comes from the definition of crossing angle $2\theta$)
\bea
\xi_{y}&=&\frac{Nr_{e}{\beta_{y}^{*}}^2}{2\pi\gamma\sigma_{x}\sigma_{y}\sigma_{z}}.
\label{Etcwe}
\eea

For the ``crabbed waist" collision, $\beta_{y}^{*}\approx \sigma_{x}/\theta$. Therefore, Eq. (\ref{Etcwe}) can be rewritten as 
\bea
\xi_{y}&\approx &\frac{Nr_{e}}{2}\frac{\beta_{y}^{*}}{\pi\gamma\sigma_{y}\sigma_{z}\theta}\approx\frac{Nr_{e}}{2\epsilon_{N}}\frac{\sigma_{x}}{\beta_{x}^{*}\sigma_{z}\theta}\propto \frac{Nr_{e}}{2\epsilon_{N}\sqrt{1+\phi^{2}}}.
\label{Ets2}
\eea
Comparing with Eq. (\ref{Ehts}), we already see that the significant decrease of $\xi_y$ with large Piwinski angle. In the ``crabbed waist" scheme, the luminosity can be enhanced by a factor of $\sigma_{z}/\beta_{y}^{*}$ as we can make $\beta_{y}^{*}\ll\sigma_{z}$.

The luminosity in ``crabbed waist" collisions is given by
\bea
\mathscr{L}&=&\frac{fN_{1}N_{2}}{4\pi\sigma_{y}\sigma_{z}\theta}\approx\frac{fN_{1}N_{2}\beta_{y}^{*}}{4\pi\sigma_{x}\sigma_{y}\sigma_{z}}\approx\frac{fN_{1}\gamma\xi_{y}}{2r_{e}\beta_{y}^{*}}.
\label{Elcw}
\eea
Note, Eq. (\ref{Elcw}) agrees with Eq. (10) in Ref. \cite{TLyons} as $\sigma_{y}\ll\sigma_{x}$. Now, for the energy of $200\ GeV$ electron-positron collider, putting the parameters in Table \ref{table}, we can get
\bea
\mathscr{L}&=&\frac{22.8mA/2\times 391000\times 0.18}{2\times 2.82\times10^{15}\,m\times 1.602\times10^{-19}C}=1.5\times10^{35}\ cm^{-2}s^{-1}.
\eea

By comparing the result from the ``crabbed waist" collision and that from ``head-on" collision, we already see these advantages of the ``crabbed waist" scheme. 
However, beamstrahlung \cite{Htuneshift, Higgsf}. puts an additional condition on the value of $N/(\sigma_x\sigma_z)$, and thus on the luminosity of high energy $e^{+}e^{-}$
colliders. It turns out that the ``crabbed waist" scheme is of marginal benefit for a 240 GeV Higgs factory circular collider where $e^+ e^- \to Z^0 h^0$. 
 The scheme is useful at the 
$Z^0$ \cite{Bogomyagkov}, where one might search for rare $Z^0 \to \tau^{\pm} e^{\mp}$ or $Z^0 \to \tau^{\pm} \mu^{\mp}$  decays\,\cite{Abada}.

\subsection{SuperKEKB in Japan}

\hspace*{\parindent}
Belle  II will have 40x the luminosity of Belle or BaBar with only a factor of 2.2 increase in beam currents, as compared to Belle\,\cite{SuperKEKB}.
The Low Energy Ring/High Energy Ring (LER/HER) energy asymmetry reduction from 3.5/8.0  to 4.0/7.0 GeV minimizes the  synchrotron radiation from the larger currents. 
Synchrotron radiation goes as beam current times the fourth power of energy, so 
the High Energy Ring only emits  $(7/8)^4$ = 0.59x as much radiation per unit of beam current in Belle II as in Belle.
Luminosity, $\mathscr{L}$,  is proportional to  the beam current, $I$, times the  vertical beam-beam tune shift, $\xi_y$, divided by $\beta^*_y$, the vertical amplitude function at the Interaction Point. The maximum allowable $\xi_y$ is set by how much the beams can be disturbed as they pass through each other.

 $$\mathscr{L} = {{1}\over{4e\,r_e}}  {{\xi_y} \over  {\beta^*_y}} \, \gamma I \eqno{(\ref{Eeeheadon})}$$

Belle II introduces a large  41.5 mrad Piwinski $e^+e^-$ beam  crossing half angle $\phi$ to allow a small $\beta^*_y$ while avoiding the hourglass effect which lowers luminosity.
The $\beta^*_y$ is reduced by a factor of 17.4  from 5.9\,mm at KEKB to an average of 0.34\,mm at SuperKEKB.
The beam overlap length is much shorter than the beam bunch length $\sigma_z$, which is hard to shorten.
Of course, introducing a large crossing angle and small $\beta^*_y$ lowers the beam-beam tune shift 
$(\xi_y \approx r_e \,N \beta_y^* / [2 \pi \, \gamma \, \sigma_y \, \sigma_z \,\phi\,])$
and hence the luminosity.  But the luminosity can be restored by reducing the vertical size, $\sigma_y$,  of
$e^+e^-$ bunches, i.e.~by using nanobeams to increase the vertical tune shift to as high a value as the storage rings will tolerate. 
If $2\,\nu_x$ or $2\,\nu_z$ or  $(\nu_x + \nu_z)$ is integral, where $\nu_x$ and $\nu_z$ are the horizontal and vertical betatron oscillation frequencies per revolution, 
resonances will occur.  Shifting a tune toward a resonance must be avoided.
The precision damping rings recently developed in the International Linear Collider R\&D program have made nanobeams possible.

\section{Proton-Proton Collider}

\hspace*{\parindent}For $e^{+}e^{-}$ colliders, the transverse beam size is  naturally damped by synchrotron radiation. However, hadrons colliders cannot enjoy fast damping due to the synchrotron radiation, at least for energies less than 10 TeV. Hence for the quest for high luminosity (smaller beam emittances) of $pp$ or $p\bar{p}$ colliders, we may either generate low emittance beams in the sources or arrange beam ``cooling" (phase space reduction, usually at low or medium energy accelerators in the injector chain), using either ``stochastic cooling" or ``electron cooling" methods \cite{hadroncollider}.

For ``round beam cross sections", the beam-beam tune shift per collision can be calculated by using Eq. (\ref{Ets}). For example, consider the shift for the 
Tevatron collider\,\cite{Tevatron2007}.
The typical numbers are $N_p = 2.5\times 10^{11}$, $\epsilon_{N,{p}}= 2.8 \pi \, mm \ mrad=3\pi \times 10^{-6}\ m$ and for the proton, $r_{p}=1.53\times 10^{-18}$ meters. 
We have

\bea
\xi_{\,\bar{p}}&=&\frac{N_p r_{p}}{4\epsilon_{N,p}}=\frac{2.5\times 10^{11}\times1.53\times 10^{-18}\,m}{4\times 2.8 \, \pi \times 10^{-6}\,m}=0.011 {\rm \, per \, IP}
\eea

For hadron colliders it is difficult to lower the emittance to raise the tune shift, but the number of hadrons per bunch might be raised even if this requires lowering the number 
of bunches
\cite{LHCoptics,OPENbeam}.  The shorter collision region may lead to vertex reconstruction pileup.

\section{Muon-Muon Collider}

\hspace*{\parindent}The $\mu^{+}\mu^{-}$ collider is attractive because the muon is a point-like particle, just like the electron but 200 times 
heavier\,\cite{Gallardo,Ankenbrandt}.   Therefore, the beamstrahlung and synchrotron radiation are not important uless energy is extremely high.
However, short lifetime of muons (around $2.2 \mu s$ in the rest frame) makes a muon collider challenging technologically. Indeed, there are still many challenges to increase the luminosity, e.g., emittance reduction\,\cite{Summers2015}, and targeting. We need also consider neutrino radiation  \cite{233km}. 

Nevertheless, it is still possible to combine design ideas of $e^{+}e^{-}$ and $pp$/$p\bar{p}$ colliders and take their advantages. If  the ``crabbed waist" scheme
allows 3x more muons per bunch, the luminosity increase 9x. This assumes that the muons are available.

For a ``round beam" muon-muon collider, the tune shift can be given by Eq. (\ref{Ets}).  As an example, let's consider the muon-muon collider parameters in Table \ref{Muontable}. The tune shift is
\bea
\xi_{y}&=&\frac{Nr_{\mu}}{4\epsilon_{N}}=\frac{2\times 10^{12}\times 2.82\times 10^{-15}\times \frac{0.511}{105.658}}{4\times 25\pi \times 10^{-6}}=0.087.
\eea

\begin{table}[ht] 
\caption{Muon Collider Parameters \cite{Mcollider} }\smallskip
\centering   
\tabcolsep=1.2mm   
\begin{tabular}{l|c c } 
\hline\hline\
$E\ (TeV)$ & 1.5 & 3\\ 
luminosity $\mathscr{L}\ (10^{34}\ cm^{-2}s^{-1})$ & 0.92 & 3.4 \\
beam-beam Tune Shift $\xi_{y}$ & $\approx 0.087$ & $\approx 0.087$ \\
number of particles per bunch $N\ (10^{12})$ & 2  & 2\\
muon transverse emittance ($\pi mm \ mrad$) & 25 & 25\\
\hline\hline     
\end{tabular} 
\label{Muontable}  
\end{table}

\section{Conclusion}

In this work, we discussed the potential advantages of the ``crabbed waist" scheme of beam-beam collisions to increase the luminosity. Some qualitative analyses are present. For $e^{+}e^{-}$ colliders, the ``crabbed waist" scheme  is possible to gain luminosity due to the significant decrease of $\beta_{y}^{*}$ (with constant $\xi_{y}$ achieved by lower emittance) 
at low energy level. For $pp$/$p\bar{p}$ and $\mu^{+}\mu^{-}$ colliders, the way to 
exploit the ``crabbed waist" to
increase the luminosity requires higher beam intensity, i.e., increasing the number of particles of each bunch. At high energy level (above the $Z^0$ mass), the ``crabbed waist" 
scheme would be of marginal  benefit because of the beamstrahlung for $e^{+}e^{-}$ colliders. Also, we should notice that though small beam sizes would bring possible high luminosity, operation of the colliders with smaller and smaller beams would also bring up many issues relevant to alignment of magnets, vibrations and long-term tunnel stability, which should be dealt with seriously. The expression of the beam-beam tune shift of $e^{+}e^{-}$ colliders is different from that of $pp/p\bar{p}$ and $\mu^{+}\mu^{-}$ colliders by a factor $1/2$ because of the beam is flat.

\end{document}